\documentstyle[aps,preprint]{revtex}
\def\H{{\cal H}}
\def\v#1{\mbox{\boldmath$#1$}}
\def\spmtot{S^{\pm}_{\mbox{tot}}}
\def\nseg{N_{\mbox{s}}}
\def\pcr{p_{\mbox{c}}} 
\begin{document}
\title{
Low Energy Properties of the Random Spin-1/2 Ferromagnetic-Antiferromagnetic
Heisenberg Chain 
}
\author{Kazuo Hida}
\address{
Department of Physics, Faculty of Science, \\ Saitama University, Urawa,
Saitama 338}
\date{\today}
\maketitle
\begin{abstract}
The low energy properties of the spin-1/2 random Heisenberg chain with
ferromagnetic and antiferromagnetic interactions are studied by means of the
density matrix renormalization group (DMRG) and real space renormalization
group (RSRG) method for finite chains. The results of the two methods are
consistent with each other. The deviation of the gap distribution from that
of the random singlet phase and the formation of the large-spin state is
observed even for relatively small systems. For a small fraction of the
ferromagnetic bond, the effect of the crossover to the random singlet phase
on the low temperature susceptibility and specific heat is discussed. The
crossover concentration of the ferromagnetic bond is estimated from the
numerical data.
e-mail: hida@riron.ged.saitama-u.ac.jp
\end{abstract}

\vspace{2mm}
\noindent

\newpage

\section{Introduction}
Recently, the physics of random quantum spin chains has been attracting the
broad interest of theoretical and experimental
studies.\cite{bl1,dgm1,dgm2,hj1,hj2,df1,wes1,wes2,furu1,furu2,kh1,hy1,f1,f2,
m1,sig1} It has been clarified that various exotic phases which are realized
in neither regular quantum systems nor classical random systems appear in
these systems.
 The interplay of quantum fluctuation and randomness is essential in
understanding the low temperature thermodynamics of these systems.

The most widely used theoretical technique for this type of problem is the
real space renormalization group (RSRG)
method.\cite{dgm1,dgm2,hj1,hj2,df1,wes1,wes2} In this approach, the {\it
distribution} of the parameters such as the bond strength or the spin
magnitude are renormalized step by step by changing the energy scale. The
ground state phases are characterized by the fixed point distribution
functions. In contrast to the RSRG method in the regular system, the RSRG
method for the distribution function is often aymptotically accurate because
of the broadness of the fixed point distribution. In the case of the random
antiferromagnetic Heisenberg chain (RAFHC),\cite{dgm1,dgm2,hj1,hj2,df1} it
is known that the ground state is the random singlet (RS) phase in which the
spins form singlets randomly not only with nearest neighbors but also with
distant partners. Unlike the RVB state, however, the spatial pattern of the
dimer covering is randomly fixed and does not fluctuate quantum mechanically. 

The RSRG study of the random Heisenberg model with both ferromagnetic (F)
and antiferromagnetic (AF) bonds (hereafter abbreviated as RFAFHC; random
ferromagnetic-antiferromagnetic Heisenberg chain) has been carried out by
Westerberg {\it et al}.\cite{wes1,wes2} Surprisingly, they predicted that
this model belongs to a different universality class from the RS phase. In
the presence of the ferromagnetic bonds, spins do not always die out but
form large effective spins of various magnitude. Thus the fixed point is
characterized by a fixed point distribution of the bond strength and spin
magnitude even if the original system consists of only spins with magnitude
1/2. This type of ground state is called the large-spin phase.

On the other hand, the present author has introduced an algorithm which
enables the application of the density matrix renormalization group (DMRG)
method \cite{white1,white2} to the random chains. \cite{kh1} This method is
applied to the RAFHC and gives results consistent with the RSRG theory.
\cite{df1,kh1} 

In the present work, the ground state and low energy properties of RFAFHC
are investigated with the help of the DMRG and RSRG methods. In the next
section, the Hamiltonian studied in this paper is presented. The numerical
results are presented in section 3. The qualitative feature of the low
temperature behavior and the crossover to the RS state in the limit of low
F-bond concentration are discussed in section 4. In the last section, our
results are summarized.  

\section{Model Hamiltonian}

The Hamiltonian of the spin-1/2 RFAFHC is defined by 
\begin{equation}
\label{eq:ham}
\H =\sum_{i=1}^{N} 2J_i\v{S}_{i}\v{S}_{i+1},\ \ \mid \v{S}_{i}\mid = 1/2,
\end{equation}
where $J_i$ takes random values of both positive and negative signs. For the
numerical calculation, we assume the following bond distribution $P_0(J_i)$,
\begin{equation}
P_0(J_i) = \left\{\begin{array}{ll}
{\displaystyle\frac{1-p}{W}} &  0 < J_{\mbox{min}} < J_i < J_{\mbox{max}}, \\
{\displaystyle\frac{p}{W}}   & -J_{\mbox{max}} < J_i < -J_{\mbox{min}} < 0, \\
0 & \mbox{otherwise},
\end{array}\right.
\end{equation}
where $W=J_{\mbox{max}}-J_{\mbox{min}}$. The absolute values of $J_i$'s are
distributed uniformly between $J_{\mbox{min}}$ and $J_{\mbox{max}}$. The
sign of $J_i$ is F and AF  with probability $p$ and $1-p$, respectively.
Hereafter we set the energy unit by $J_{\mbox{max}}=1$. 

Furusaki {\it et al}.\cite{furu1,furu2} investigated the finite temperature
properties of the random exchange Heisenberg model with the $\pm J$ bond
distribution,
\begin{equation}
P_0(J_i) = p\delta(J_i+J)+(1-p)\delta(J_i-J).
\end{equation}
This model is called $\pm J$ Heisenberg chain (or $\pm J$ HC) hereafter. The
present model reduces to  $\pm J$ HC in the limit $J_{\mbox{max}}
\rightarrow J_{\mbox{min}}+0$. However there is a significant difference
between the present model and $\pm J$ HC in the limit $p \rightarrow 0$.
Namely, in this limit, $\pm J$ HC reduces to the uniform spin-1/2 Heisenberg
chain while the present model reduces to the RAFHC as long as
$J_{\mbox{max}} \neq J_{\mbox{min}}$. Therefore the present model is
suitable for investigating the crossover of the low energy properties of the
RFAFHC to those of RAFHC as the concentration of the F-bond tends to 0.
\section{Numerical Results}
The algorithm of the DMRG calculation is essentially the same as that
introduced in ref. \cite{kh1}. In the present model, however, even the
ground state is not always a singlet. Working in the subspace with
$S^z_{\mbox{tot}}=0$, the obtained target states generally have the finite
total spin $S_{\mbox{tot}}$. In each step of renormalization, 2 spins are
added. To obtain the low lying states of the superblock with $N+2$ spins
with $S^z_{\mbox{tot}}=0$, the important states of the left and right blocks
of the $N$ spin chain with  $S^z_{\mbox{tot}}=\pm 1$ are taken into account,
because the two additional spins can change $S^z_{\mbox{tot}}$ by $\pm 1$.
The states with $S^z_{\mbox{tot}} = \pm 1$ are generated by the application
of the ascending and decending operators of total spin $\spmtot$ to the
target state with $S^z_{\mbox{tot}}=0$. This means that we have to keep
three times the number of states compared to the case of the purely AF
chain. In our calculation, we kept 140 states in each iteration step. 

To compare the DMRG results with the RSRG results, we have also performed
the RSRG calculation\cite{wes1,wes2} for the finite systems. In contrast to
Westerberg {\it et al}.,\cite{wes1,wes2} we do not add extra spins after
each decimation. Therefore the effective length of the chain decreases step
by step. The energy gap of the last step effective Hamiltonian is identified
as the energy gap of the whole system. This procedure is performed for many
samples and the gap distribution is calculated.

 In the RS phase of RAFHC, the distribution of the logarithm of the energy
gap $\Delta$ of the chain of length $N$ is scaled as
$(\ln\Delta)/N^{1/2}$.\cite{df1} This implies that the average of
$\ln\Delta$ scales with $N^{1/2}$. This property is also confirmed to hold
even for relatively small systems by the DMRG method.\cite{kh1} On the other
hand, according to Westerberg {\it et al}.,\cite{wes1,wes2} the energy gap
is scaled by the power of $N$ in RFAFHC as $\Delta \sim N^{-1/\delta}$ with
$\delta \simeq 0.44$. In this case, the average of $\ln\Delta$ should scale
with $\ln N$. This difference in the energy spectrum is reflected in the
specific heat and magnetic susceptibility at low
temperatures.\cite{df1,wes1,wes2}

For the DMRG calculation, the number of samples ranges from 120 to 180 and
the maximum system size is around 60. For the RSRG calculation, the number
of samples is 2500. We have studied the cases of $p = 0, 1/5,$ $ 1/3$ and
$2/3$ with  $J_{\mbox{min}}=0.5$ and $J_{\mbox{max}}=1$. Figure \ref{fig1}
shows the system size dependence of the average $<\ln \Delta>$ plotted
against $N^{1/2}$, which should be a straight line with a finite gradient
for RAFHC, as in the case of $p =0$. For $p = 2/3$, the deviation is
significant even for the systems with $N \leq 60$ studied here. It should be
noted that the results of RSRG and DMRG are consistent with each other as
far as the $N$-dependence is concerned, although there is discrepancy in the
absolute values for large $p$. The fact that the RSRG result concides
semi-quantitatively with the DMRG result indicates the reliability of the
RSRG results for larger systems, because the RSRG method becomes more and
more accurate as the gap distribution broadens. Unfortunately, it is not
possible to obtain the value of the exponent $\delta \simeq 0.44$ obtained
by Westerberg {\it et al}.\cite{wes1,wes2} within the DMRG data. This means
that the approach to the fixed point distribution is much slower than in the
case of RAFHC for which the fixed point distribution is already reached for
$N \leq 60$ within the DMRG data. Actually, Westerberg {\it et
al}.\cite{wes1,wes2} used the chains of $10^5$ to $10^6$ sites to reach the
fixed point distribution. 

 Figure \ref{fig2} shows the system size dependence of the average of the
square of the total spin magnitude $<S_{\mbox{tot}}^2>$. It is clearly seen
that $<S_{\mbox{tot}}^2>$ is proportional to $N$ which is the specific
feature of the large-spin phase of Westerberg {\it et al}.\cite{wes1,wes2}
Here again the DMRG and RSRG data are consistent with each other. The ratio
$c= <S_{\mbox{tot}}^2>/N$ is plotted against $p$ in Fig. \ref{fig3}. The
solid line shows the fixed point relation $c = p/4(1-p)$ for
$S=1/2$.\cite{wes1,wes2} Thus our data show that the ground state is already
clearly distinct from the RS ground state even for relatively small systems,
although the approach to the true fixed point is extremely slow.

\section{Crossover of the Low Energy Behavior}

The extremely slow convergence of renormalization flow prevents us from
direct numerical investigation of the ground state of the present model
especially for small $p$. Nevertheless, the crossover to the RS state in the
limit $p \rightarrow 0$ can be speculated by combining our observations for
the small systems and the analytical results obtained so far.

 For small $p$, most of the spin pairs are connected by the random AF-bonds.
In terms of the RSRG scheme, these spins are killed in the early stage of
decimation leading to finite segments of the random singlet phase (RS
segments) connected by the F-bonds. The energy gap distribution in the RS
segment is given by the RS fixed point distribution function with finite
energy cut-off $\Omega$ as\cite{wes1,wes2}
\begin{equation}
\label{fix}
P_{\mbox{RS}}(\Delta; \Omega) = \frac{\alpha}{\Omega}\left[\frac{\Omega}
{\Delta}\right]^{1-\alpha}\theta(\Omega-\Delta),
\end{equation}
with $\alpha = 1/\ln \Omega^{-1}$.The cut-off $\Omega$ and the exponent
$\alpha$ is related to the size of the segment $\nseg$ as\cite{df1,kh1}
\begin{equation}
\label{alp}
 \ln(1/\Omega(\nseg))=\alpha(\nseg)^{-1} \simeq C_0 + C_1 \nseg^{1/2}, 
\end{equation}
where $C_0$ and $C_1$ are numerical constants of the order of unity. The
typical length of the RS segment is $1/p$ and its energy scale is
$\Omega(1/p)$. The spins in the even-length RS segment can form complete
random singlet states within each segment, while in each odd-length segment
a spin-1/2 degree of freedom remains. These spin-1/2 degrees of freedom are
coupled weakly via even-length RS segments with the effective bond strength
of the order of $\Omega(1/p)$.

Based on the observation above, we can speculate the finite temperature
behavior for small $p$,  as schematically summarized in Fig. \ref{fig4}. In
the intermediate temperature regime between the typical exchange energy of
the original Hamiltonian and the cut-off energy of the typical RS segment
$\Omega(1/p)$, the temperature $T$ determines the effective energy scale and
the cut-off $\Omega(1/p)$ is irrelevant. Therefore the contribution to the
susceptibility from the RS segments takes the usual RS form\cite{hj1,df1}
and gives the dominant contribution as
\begin{equation}
\chi \sim  \frac{\mu_{\mbox{B}}^2}{T\mid\ln^2 T\mid}.
\end{equation}
It should be noted that the spin-1/2 degrees of freedom in the odd-length AF
segments gives the Curie-type contribution to the susceptibilty $\sim p/T$
because their number is proportional to $p$. Nevertheless, the RS
contribution is always dominant in this regime, because $\ln^{-2}T >> p$ as
far as $T >> \Omega(1/p)$. The specific heat is similarly given by the RS
form:\cite{hj2}
\begin{equation}
C \sim  \frac{1}{\mid\ln^3T \mid}.
\end{equation}
This regime is denoted by I in Fig. \ref{fig4}.

As the temperature becomes lower than $\Omega(1/p)$, distribution
(\ref{fix}) with cut-off $\Omega(1/p)$ gives the effective gap distribution
in RS segments. At the same time, the spin-1/2 degrees of freedom in the
odd-length segments start to be correlated. If the distribution
$P_{\mbox{RS}}(\Delta; \Omega(1/p))$ is less singular than the universal
fixed point distribution $P_{\mbox{U}}(\Delta) \sim \Delta^{-y_c}$ with $y_c
\sim 0.7$ obtained by Westerberg {\it et al}.,\cite{wes1,wes2} the
renormalization flow will be attracted to this universal fixed point. This
regime is denoted by regime II in Fig. \ref{fig4}. The susceptibility and
the specific heat are given by,
\begin{equation}
\label{chi}
\chi \sim \frac{\mu_{\mbox{B}}^2p}{12T(1-p)},
\end{equation}
\begin{equation}
C \sim T^{\delta}\mid\ln T\mid \ \ \ \mbox{with} \ \ \ \delta \simeq 0.44,
\end{equation}
following refs. \cite{wes1,wes2}. 

On the other hand, if the distribution  $P_{\mbox{RS}}(\Delta; \Omega(1/p))$
is more singular than the universal fixed point distribution, the low energy
physics is governed by the finite size RS distribution
$P_{\mbox{RS}}(\Delta; \Omega(1/p))$ for the finite segments. Therefore, the
specific heat is dominated by the finite size RS contribution.
\begin{equation}
C \sim T^{\alpha(1/p)}.
\end{equation}
similarly to the case of random dimer phase.\cite{hy1} However, the
odd-length RS segments still involve the spin-1/2 degrees of freedom which
contribute to the susceptibility. Considering that the effective magnitude
of the cluster spin scales with the square of the cluster size even for the
chains with singular initial bond distribution close to the RS fixed point
distribution (chain E of ref. \cite{wes2}), these spin-1/2 degrees of
freedom would also contribute to Curie law susceptibility (\ref{chi}) in the
similar way as in the universal case.  This regime is denoted by III in Fig.
\ref{fig4}. 
Physically speaking, the spin degrees of freedom in the RS segments die out
as the temperature is lowered and therefore the contribution to the
susceptibility is less singular than the Curie law. However, energetically,
there remain arbitrarily weak singlet pairs in the RS segments which
dominate the low temperature specific heat in this regime.

As $p \rightarrow 0$, the lower limit of the intermediate regime
$\Omega(1/p)$ tends to 0 and the RS behavior $\chi \sim
\mu_{\mbox{B}}^2/(T\ln^2T)$ is recovered down to $T=0$ in this limit. This
is in contrast to the case of $\pm J$ HC studied by Furasaki {\it et
al}.,\cite{furu1,furu2} for which the thermodynamics at the intermediate
temperature is described by the assembly of F and AF segments and approaches
the uniform AF chain as $p \rightarrow 0$.

Ths critical concentration $\pcr$ between regimes II and III is determined
by $y_c = 1-\alpha(1/\pcr)$. The numerical value of $\pcr$ can be estimated
from the DMRG data for $p=0$ (RAFHC) as follows. The $N$-dependence of
$\alpha(N)$ can be deduced from the formula $\mid<\ln \Delta>_{p=0}\mid/2 =
\alpha(N)^{-1}$ verified by integrating (\ref{fix}). Fitting the DMRG data
for $p=0$ to (\ref{alp}), we find $C_1 \simeq 0.3$ (Fig. \ref{fig1}) while
$C_0$ depends on the choice of the energy unit by definition. Although the
energy unit was fixed by $J_{\mbox{max}} =1$ in the beginning, here it is
more appropriate to redefine the energy unit so that the relation $-<\ln
\Delta>/2=\sigma \equiv <(\ln\Delta-<\ln \Delta>)^2>^{1/2}$, which is valid
for the distribution (\ref{fix}), holds for large enough $N$. This leads to
$C_0 \simeq -0.78$ and $\pcr \simeq 0.0053$. This procedure is better than
the direct fit to $\sigma$ because the convergence of $\sigma$ to the fixed
point value is slower than $<\ln \Delta>$. It would be quite difficult to
access the region $p < \pcr$ by the DMRG method, because systems much larger
than 190 ($\sim 1/0.0053$) would be required for this purpose.

\section{Summary}

The low energy properties of the random ferromagnetic-antiferromagnetic
quantum Heisenberg chain with spin-1/2 is studied by means of the DMRG and
RSRG methods. It is demonstrated that the RSRG scheme gives results
consistent with those of DMRG even for relatively small systems, which
confirms the accuracy of the RSRG scheme for larger systems. It is shown
that the distribution of the logarithm of the gap deviates from that of the
random singlet phase and $<S_{\mbox{tot}}^2>$ grows with the system size
indicating the transition to the large-spin phase. However, the size of the
system tractable by the DMRG is too small to approach the fixed point.
Based on the numerical and analytical results obtained so far, the physical
picture of the crossover of the low energy behavior to the random singlet
phase is discussed. The critical concentration of the F-bonds is estimated
from the numerical data.

The author thanks A. Furusaki for stimulating communication. The numerical
calculations were performed using FACOM VPP500 at the Supercomputer Center,
Institute for Solid State Physics and HITAC S820/15 at the Information
Processing Center, Saitama University. This work is supported by a
Grant-in-Aid for Scientific Research from the Ministry of Education,
Science, Sports and Culture.

\begin{figure}
\caption{
The system size dependence of the average $ < \ln \Delta > $ plotted against
$N^{1/2}$ for $p =0, 1/5, 1/3$ and  2/3 with $J_{\mbox{min}}=0.5$ and
$J_{\mbox{max}}=1.0$. The DMRG data for $p=0$ are shown by $\times$. For $p
\neq 0$, filled symbols are DMRG data and open symbols are RSRG data.
}
\label{fig1} 
\end{figure}
\begin{figure}
\caption{
The system size dependence of $<S_{\mbox{tot}}^2>$ for $p = 1/5, 1/3$ 
and  2/3  with $J_{\mbox{min}}=0.5$ and $J_{\mbox{max}}=1.0$. Filled symbols
and solid lines are DMRG data and open symbols and broken lines are RSRG
data. The lines are the least squares fits to the data.
}
\label{fig2}
\end{figure}
\begin{figure}
\caption{
The $p$-dependence of $c=<S_{\mbox{tot}}^2/N>$ estimated from DMRG data 
($\bullet$), RMRG data ($\circ$) and fixed point value $p/4(1-p)$ by 
Westerberg and coworkers.[7,8](solid line). 
}
\label{fig3}
\end{figure}
\begin{figure}
\caption{
Schematical low temperature behavior of RFAFHC for small $p$. 
All lines are crossover lines. At $T=0$, the ground state corresponds to 
the universal fixed point for $p > p_{\mbox{c}}$ and to the non-universal
one for
 $p < p_{\mbox{c}}$.
}
\label{fig4}
\end{figure}
\end{document}